\documentclass[twoside,12pt]{article}
\usepackage{graphicx}

\begin{document}

\centerline{\Large \bf Ising Ferromagnet:}\medskip
\centerline{\Large \bf  Zero-Temperature Dynamic Evolution}

\bigskip

P.M.C. de Oliveira$^{1,*}$, C.M. Newman$^2$, V. Sidoravicious$^3$, and
D.L. Stein$^{2,4}$

\bigskip

\noindent 1) Instituto de F\'\i sica, Universidade Federal Fluminense\par

av. Litor\^anea s/n, Boa Viagem, Niter\'oi RJ, 24210-340 Brazil

\noindent 2) Courant Institute of Mathematical Sciences, New York University,\par

251 Mercer St, New York, NY 10012 USA\par

\noindent 3) Instituto de Matem\'atica Pura e Aplicada\par

Estrada D. Castorina 110, Rio de Janeiro RJ, 22460-320 Brazil\par

\noindent 4) Department of Physics, New York University,\par

New York, NY 10003 USA\par

\medskip

\noindent $*$ e-mail: pmco@if.uff.br

\medskip

PACS numbers: 02.70.-c, 05.10.Ln, 64.60.Ak, 05.70.Jk

\bigskip

\begin{abstract}

	The dynamic evolution at zero temperature of a uniform Ising
ferromagnet on a square lattice is followed by Monte Carlo computer
simulations. The system always eventually reaches a final, absorbing state,
which sometimes coincides with a ground state (all spins parallel), and
sometimes does not (parallel stripes of spins $up$ and $down$). We initiate
here the numerical study of ``Chaotic Time Dependence'' (CTD) by seeing how
much information about the final state is predictable from the randomly
generated quenched initial state. CTD was originally proposed to explain how
nonequilibrium spin glasses could manifest equilibrium pure state structure,
but in simpler systems such as homogeneous ferromagnets it is closely related
to long-term predictability and our results suggest that CTD might indeed occur
in the infinite volume limit. \end{abstract}

\bigskip\bigskip

\section{Introduction}

	We consider the Ising model on an $L \times L$ square lattice with
periodic boundary conditions. Each site carries a spin either $up$ or
$down$, i.e. $S_{ij} = \pm 1$, $i,j = 1$, 2 $\dots$ $L$. A pair of
neighboring sites has unit energy if the two spin orientations are
antiparallel; parallel spins have no energy. The total energy thus ranges
from $E = 0$ to $2L^2$, with the lowest possible energy $E = 0$
corresponding to either of the two ground states with all spins either up or
down.

	A randomly chosen state of this system is stored into the computer
memory, and the following dynamic rule is applied to it. A site is chosen at
random, and its spin is a candidate to be flipped. If the energy decreases
as a result of this flip, then we perform it. Energy increases are not
accepted: the chosen spin keeps its current state. If the energy would be
unchanged, then the flip is performed with probability 1/2.  This procedure
is then repeated for another randomly chosen site, and so on.

	Physically, this problem corresponds to a sudden quenching from
infinite to zero temperature. It was previously studied by measuring the
ordinary magnetization, for instance in references
\cite{NS99,OP02,Spirin02,OPT04,SS05}. Here we consider it from a completely
different point of view: our interest is to study the influence of the
starting state on a configuration at a later time $t$.

	For each starting state, we perform $D$ independent runs, each
corresponding to a different realization of the dynamics (i.e., a different,
and also randomly chosen, chronological order of spins selected to be
flipped along with a different coin toss for each zero-energy flip
encountered). Each step in the time $t$ corresponds to $L^2$ flip attempts,
i.e. a whole-lattice sweep, on average. The {\bf local} quantity

\begin{equation}
<S_{ij}>\,\, =\,\, {1\over D}\,\, \sum_{d=1}^D\,\, S_{ij}
\end{equation}

\noindent is calculated at each site $ij\,$, for $t = 0$, 1, 2 $\dots$. 
Furthermore, for each $t$, the {\bf global} averages

\begin{equation}
Q(t)\,\, =\,\, {1\over L^2}\,\, \sum_{ij}\,\, <S_{ij}>
\end{equation}

\noindent and

\begin{equation}
R(t)\,\, =\,\, {1\over L^2}\,\, \sum_{ij}\,\, <S_{ij}>^2
\end{equation}

\noindent are determined, where the sums run over all $L^2$ sites.

\begin{figure}[!ht]
\begin{center}
\includegraphics[angle=-90,scale=0.45]{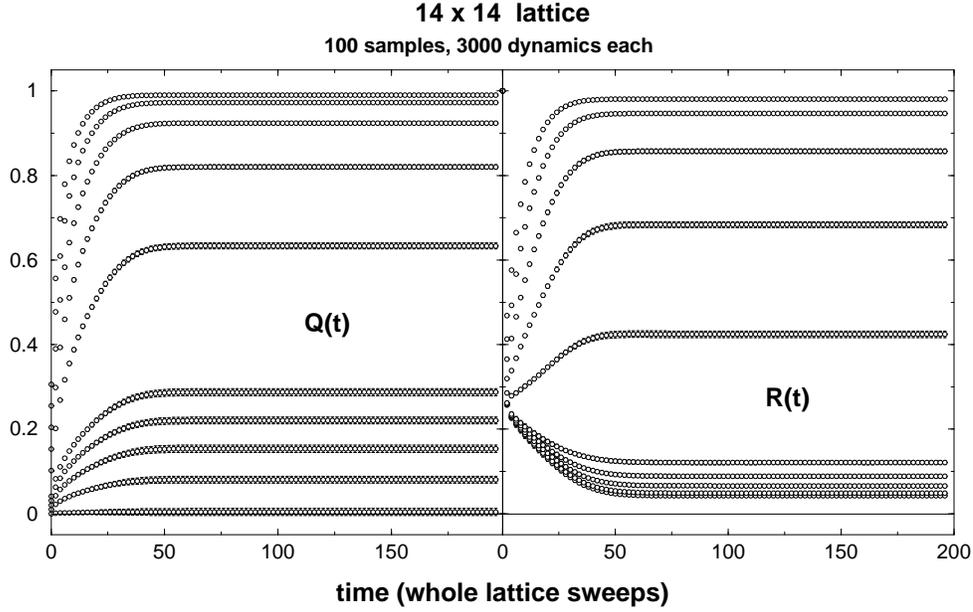}
\end{center}

\caption{Time functions $Q(t)$ and $R(t)$, Eqs.~(2) and (3), averaged over
100 different initial states. Each curve corresponds to a different starting
magnetization $m$: exactly $0$ (bottom curves in both sides), $0.01$,
$0.02$, $0.03$, $0.04$, $0.10$, $0.15$, $0.20$, $0.25$ and $0.30$ (from
bottom to top). At the beginning, $Q(t=0) = m$ and $R(t=0) = 1$ in all
cases. The error bars lie mostly within the symbols, except for some cases.
Note that only for the symmetric case $m = 0$ one gets $Q(t) = 0$ within the
error bars (left).  On the other hand, $R(t)$ does not vanish even for this
same symmetric case (right). Other lattice sizes, both smaller and larger
than $L=14$, follow the same behavior.}
 \end{figure}

	In some sense, our approach is the opposite of the process called
``damage spreading'', where two slightly different initial states are followed
exactly by the same dynamic rule, including the same sequence of spins to be
flipped and any other internal or external contingency. Here, we are interested
in the effect on the final state of different contingencies occurring during
the time evolution, i.e., different chronological orders of the spins to be
flipped and different coin tosses for deciding zero-energy flips. Starting from
the {\it same} initial configuration, equations (1), (2) and (3) compare $D$
parallel, independent evolutions of the same initial state.

	Figure~1 shows the time dependences of the global averages for $S =
100$ samples, and $D = 3000$ different dynamics each. Each sample corresponds
to a new starting configuration which is randomly chosen within a fixed value
for the magnetization at $t = 0$:

 \begin{equation}
 m\,\, =\,\, {1\over L^2}\,\, \sum_{ij}\,\, S_{ij}\,\,\,\,\,\,\,\, {\rm
for}\,\,\,\, t = 0\,\,\,\, . 
 \end{equation}

	In order to prepare the starting state, we choose exactly $(1+m)L^2/2$
sites with spins $up$, the remainder $(1-m)L^2/2$ sites with spins $down$.  Our
program stores one spin per bit along a 32-bit computer word, corresponding to
32 different starting states processed at once. For that, we profit from the
fast bitwise operations, using techniques described in~\cite{Book91}.  As a
check, we performed also the same simulations starting from completely random
configurations, instead of classifying them according to the magnetization. The
results (not shown) are similar to those obtained from $m = 0$, but with much
larger fluctuations.

	One important feature exhibited in Figure~1 is that any small non-zero
starting magnetization is enough to break the symmetry: for large enough times,
the system saturates on a non-vanishing value for $Q(t)$, {\it always larger}
than the starting magnetization itself. This is true even for larger lattices
(not shown), for which one can better approach the limit $m \to 0$. 

	As we discuss more fully in the next section when we introduce the
notion of ``Chaotic Time Dependence''~\cite{NSold}, a basic issue we wish to
explore concerns the whole set of $D$ distinct dynamical histories starting
from the same initial state. In principle, this could be studied by considering
$Q(t)$, as $t\to\infty$, without any averaging over $S$ samples (since such
averaging would give a quantity essentially the same as the average
magnetization) by seeing whether $Q(\infty)$ was nonzero for a nonnegligible
fraction of starting states. If somehow all these $D$ distinct histories
diverge from each other, how much do they keep in common due to their common
starting point? This is the question we are interested in. For initial
magnetizations $m \ne 0$, the left part of Figure~1 provides a clear answer.
For $m = 0$, instead, we look at the quantity $R(t\to\infty)$ which we can
average over $S$ samples and ask whether it stays nonzero as system size
increases.

	Based on the observation that any non-zero starting magnetization is
enough to break the $up/down$ symmetry, we conclude that the most interesting
case is the initially symmetric situation, i.e. $m = 0$ exactly. Thus,
hereafter we will treat only this case, fixing attention on the behavior of
$R(t)$.

	The text is divided into two more sections: first the description of 
our simulations and the presentation of the results, then our conclusions.

\section{Description and Results}

	A first important observation concerns the absorbing state, i.e.
the final distribution of all spins from which no more changes are
possible. Before the system reaches this situation, we call it {\it
alive}, after, it is {\it dead}. These terms apply to the whole lattice,
not to each spin: the system is {\it dead} when no energy decrease or
tie can be achieved by flipping {\it any} of its spins. As noted
earlier, there are only two possible ground states each with energy $E =
0$, with all spins either up or down. Both states are clearly absorbing.
Our simultations found that in roughly 2/3 of the realizations the
system becomes eventually dead in one of these two ground states: we
label these realizations with the symbol {\it GS}. However, within the
remainder 1/3 of the realizations, the system becomes trapped into other
absorbing states with $E > 0$. The common example is a striped
configuration, with alternating stripes of {\it up\/} and {\it down\/}
spins, whose widths are larger than one layer. Clearly, this situation
does not allow any further change, and the system becomes dead as soon
as it is reached: we label these cases with the symbol {\it ST}. All of
these findings are in agreement with earlier
studies~\cite{Spirin02,SS05}. Figure~2 shows two typical countings of GS
{\it versus\/} ST situations. Note that the approximate balance 2/3
against 1/3 does not depend on the lattice size: striped configurations
always appear, independently of the lattice size.

\begin{figure}[!ht]

\begin{center}
\includegraphics[angle=-90,scale=0.45]{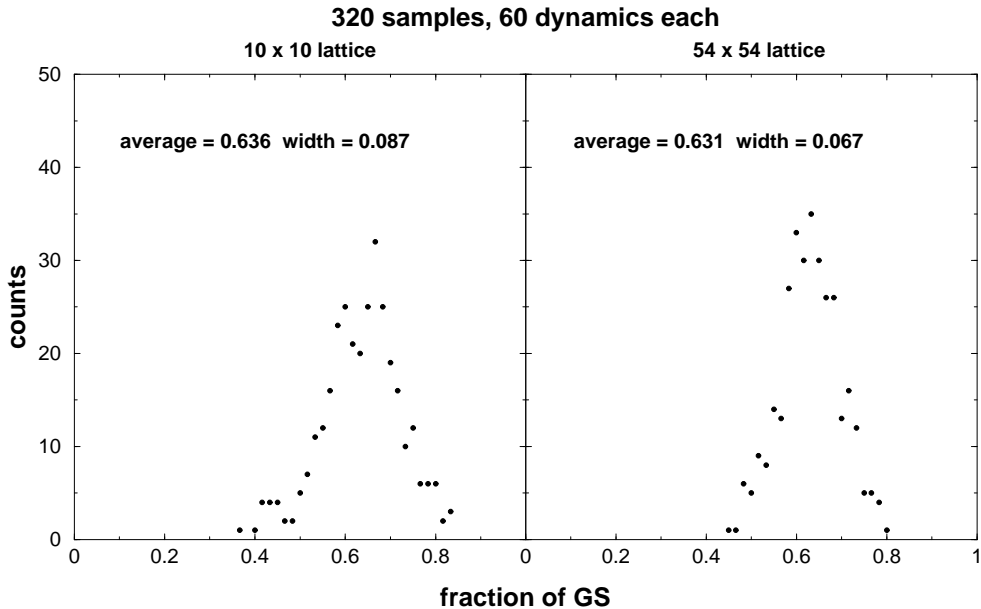}
\end{center}

\caption{Statistics of GS realizations, i.e. those for which the absorbing
state is a ground state (see first paragraph, Sect.~2). From $D$ (in this
case 60) dynamics starting from the same initial state, we record the
fraction of GS realizations obtained. The whole procedure is repeated $S$
(in this case 320) times, from which the histogram is constructed.}

\end{figure}

	We discard ST situations, keeping only GS in our statistics.
Striped configurations appear as a consequence of the finite lattice
size. In an infinite lattice, domains of up or down neighboring spins
grow forever; there is zero probability (with respect to either initial
configuration or dynamical realization) of the system evolving towards a
`striped', or domain wall, state as $t\to\infty$~\cite{NNS00}. This
means that {\it any\/} finite region eventually consists of only a
single domain (equivalently, after some fixed finite time its spin
configuration is GS). Therefore, if our simulations are to provide
insights into the infinite lattice situation, it is proper to consider
only GS realizations. All other possibilities are consequences of finite
size and are thus discarded.

\begin{figure}[!ht]

\begin{center}
\includegraphics[angle=-90,scale=0.45]{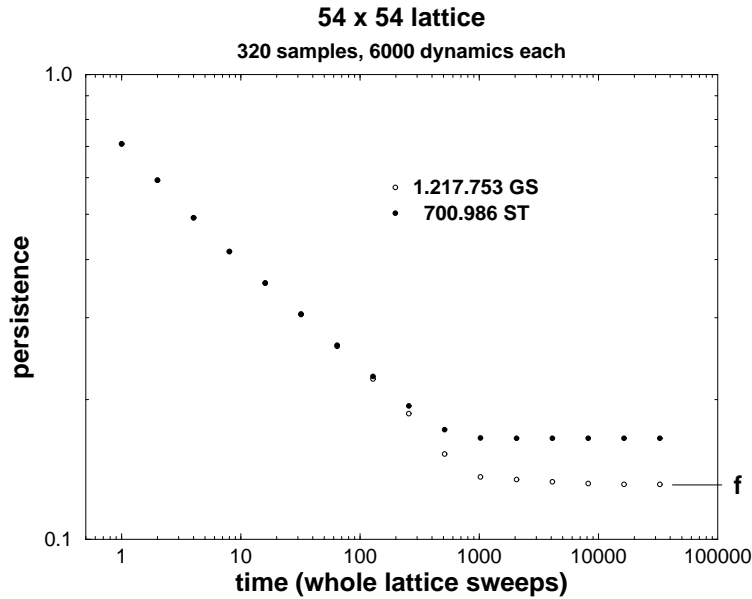}
\end{center}

\caption{The fraction of never flipped spins (persistence) decays with time
as $p \propto t^{-\theta}$. Curves with averages over only GS or only ST
realizations are shown. The critical exponent we obtained from GS statistics
is $\theta \approx 0.24$ (the same also for other values of $L$ and $D$), in
good agreement with earlier studies \cite{Derrida}. Also, the final
saturated fraction $f$ scales with the lattice size as $f \propto
L^{-\beta}$, with $\beta \approx 0.47$ extracted from our data for $L = 10$,
14, 22, 30, 38 and 54.}

\end{figure}

	There is also a `practical' reason for discarding runs terminating
in a striped configuration: they are boundary-condition-dependent. Of
course, when starting a run, one is not able to predict whether it will
evolve into a GS or ST configuration, since the outcome can {\it a priori\/}
depend on the dynamical realization. We therefore run each dynamical
realization twice. First, we note simply whether for that particular
dynamics the system reaches a GS or a ST configuration. By doing this we can
perform separate statistics for GS and ST outcomes. As an example, we
analyze a quantity defined by Derrida~\cite{Derrida}: the fraction of `never
flipped' spins as a function of time. This quantity is called {\it
persistence\/}, and was shown to display critical behavior, i.e.,
\begin{equation} p\,\, \propto\,\, t^{-\theta}\,\,\,\,\,\,\,\, ,
\end{equation} decaying as a power-law whose critical exponent $\theta$
obeys some universality properties \cite{Derrida}. It is shown in Figure~3.
Indeed, the plot obtained from only GS realizations saturates later than the
corresponding ST plot, for the same finite lattice size, allowing a better
determination of the critical behavior, a practical advantage.

	As a byproduct, we also find the further finite-size-scaling relation

\begin{equation}
 f(L)\,\, \propto\,\, L^{-\beta}\,\,\,\,\,\,\,\, {\rm
 with}\,\,\,\, \beta \approx 0.47\,\,\,\,\,\,\,\, ,
\end{equation}

\noindent for the asymptotic saturated persistence $f$, which does not
depend on the number $D$ of dynamics.

\begin{figure}[!ht]

\begin{center}
\includegraphics[angle=-90,scale=0.45]{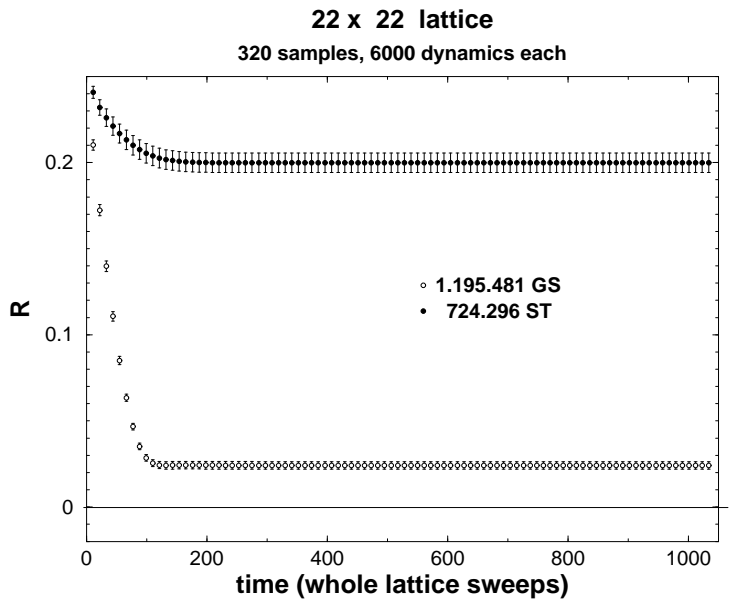}
\end{center}
\caption{Quantity $R$ as a function of time, averaged over GS (bottom) and ST
(top) situations, separately.}

\end{figure}

\begin{figure}[!ht]

\begin{center}
\includegraphics[angle=-90,scale=0.45]{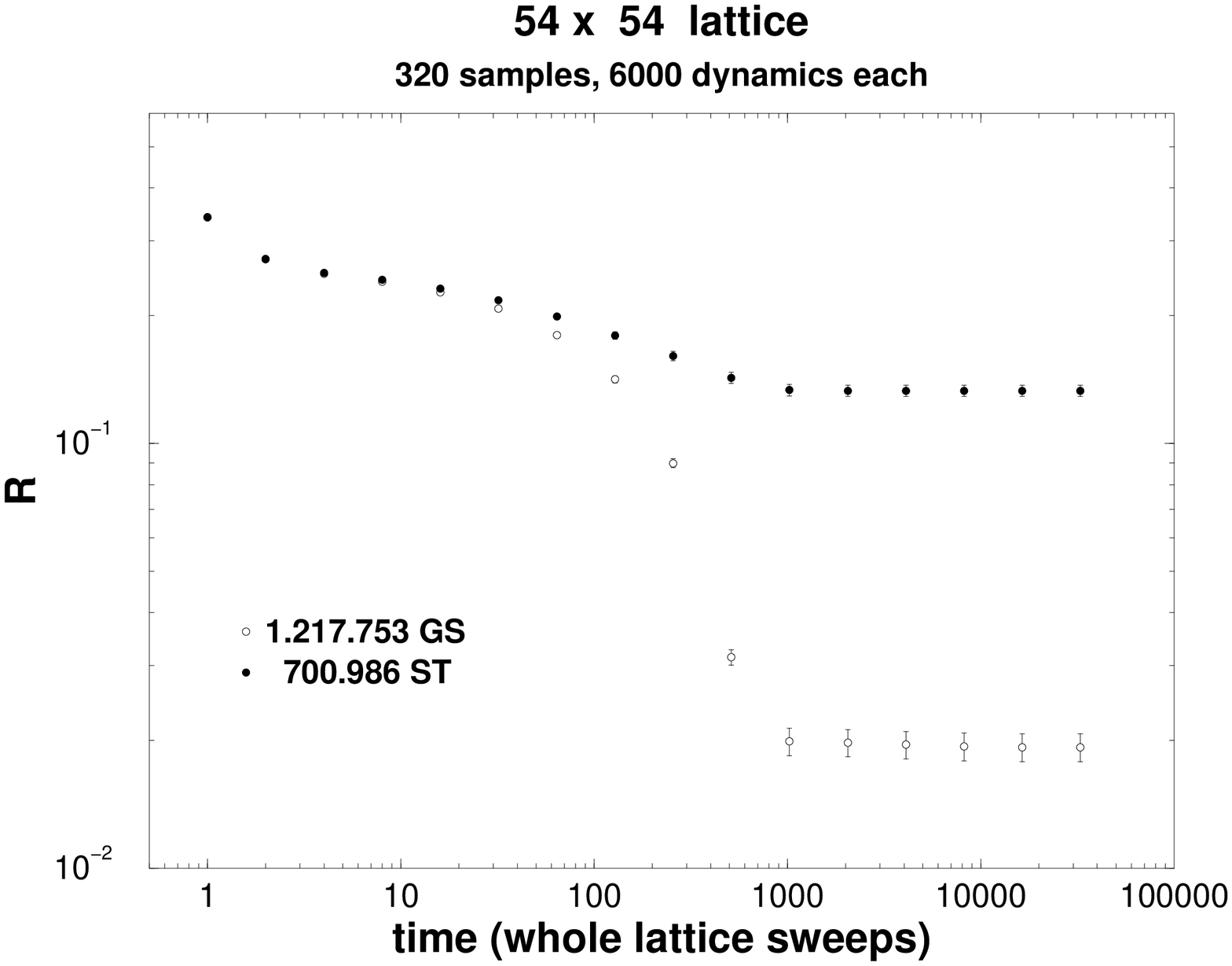}
\end{center}

\caption{The same as Figure~4, for a larger lattice and using a logarithmic
scale.}

\end{figure}

	Figure~4 shows another example, now for the quantity $R$ (cf.~Eq.~(3)).
It saturates in a much smaller value for GS than for ST. This is true also for
other lattice sizes, both smaller and larger than $L=22$. Note also the smaller
fluctuations (error bars) obtained for GS. Figure~5 shows the same behavior,
for a larger lattice, now in logarithmic scale.

\begin{figure}[!ht]

\begin{center}
\includegraphics[angle=-90,scale=0.45]{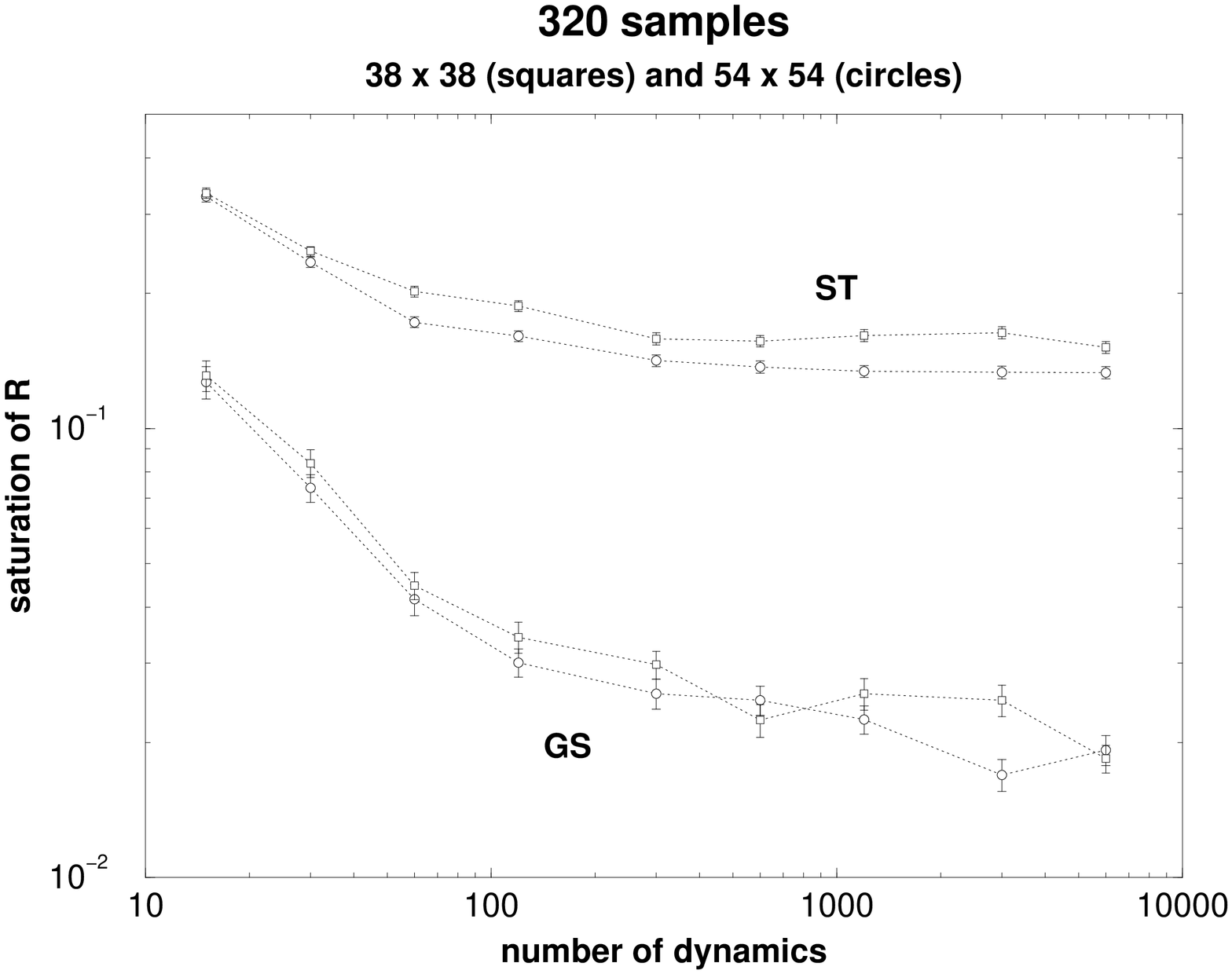}
\end{center}

\caption{Large-time asymptotic values, $R(t\to\infty)$, as a function of the
number $D$ of dynamical runs, for $L=38$ (squares) and $54$ (circles).} 

\end{figure}

	Contrary to persistence, the asymptotic value $R(t\to\infty)$ does
depend on the number $D$ of dynamics. Figure~6 illustrates the distinct
behaviors of $R(t\to\infty)$, for GS and ST realizations, as the number $D$
of dynamical runs increases. For GS the function $R(t)$ is considerably
smaller for large enough times.

	Let us denote by $R_L^{GS}$ the $D \to \infty$, $t\to\infty$ value
of $R(t)$, restricted to GS realizations (but otherwise averaged over all
initial states). This quantity provides a measure of the information about
the final state already contained in a typical, randomly chosen, initial
state. It cannot vanish for a fixed finite size $L$ because there are some
GS realizations that definitely determine the final state, independent of
dynamical realization. Figure~6 does not show significant size dependence of
$R(t)$ at large $t$, and hence suggests the possibility that $R_L^{GS}$ may
not tend to zero as $L \to \infty$. This in turn suggests that the
phenomenon of ``Chaotic Time Dependence'' (CTD)~\cite{NSold} might be
occurring.

	CTD concerns the large-time predictability of an infinite system
based on the randomly generated initial state and not dependent on the
realization of the dynamics. CTD means that $<S_{11}>$ at time $t$, averaged
over all the dynamics, in the limit $L \to \infty$, does {\it not\/} tend to
zero as $t \to \infty$ (and thus forever oscillates between positive and
negative values) for typical randomly generated initial states. We note that
in principle, CTD could occur even without the nonvanishing of $R_{L \to
\infty}^{GS}$ since CTD does not involve any restriction of initial states
(to GS) or any averaging over initial states (or equivalently over sites in
the lattice). On the other hand, it seems clear that CTD should occur if
indeed $R_{L \to \infty}^{GS} \neq 0$.

\begin{figure}[!ht]

\begin{center}
\includegraphics[angle=-90,scale=0.45]{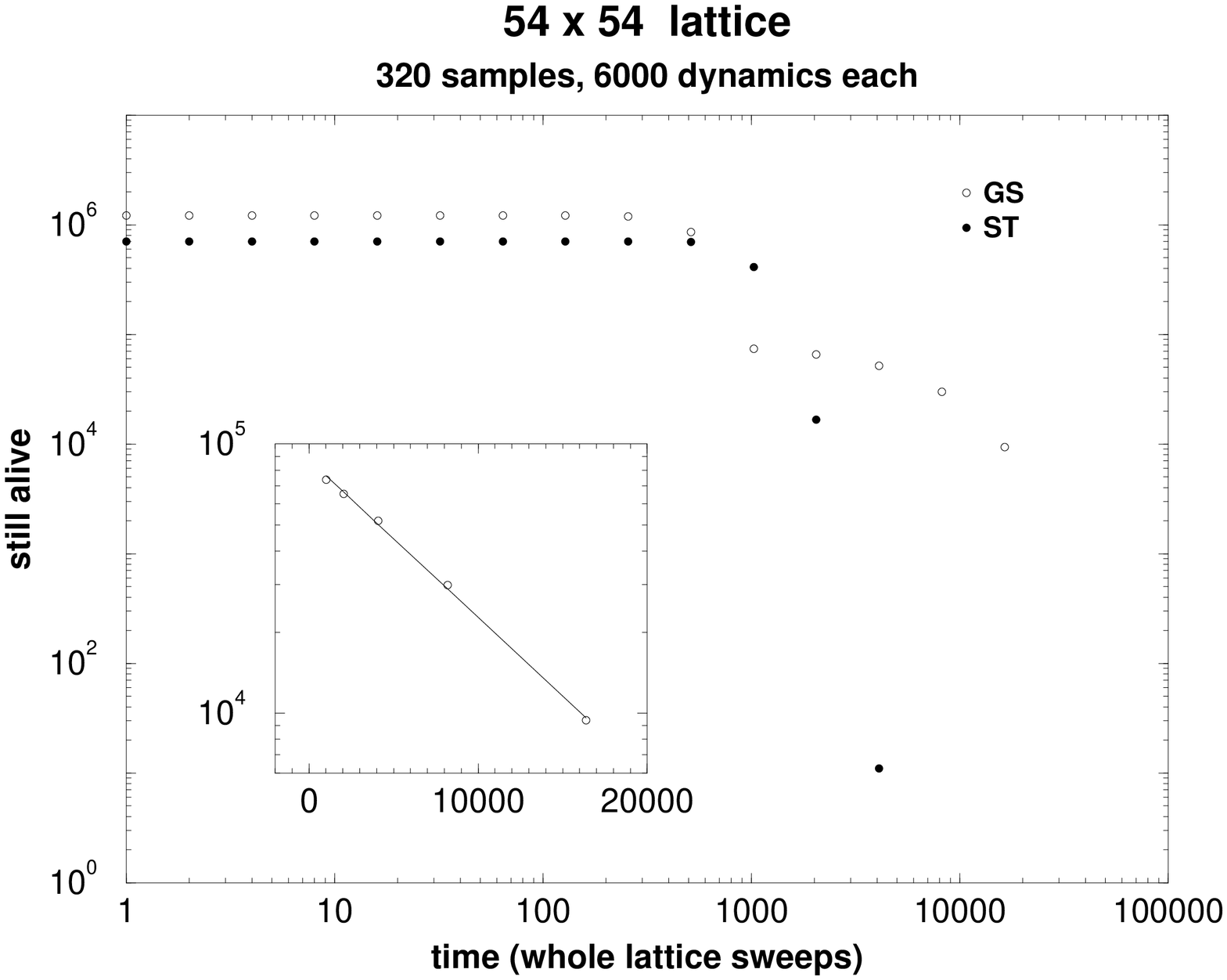}
\end{center}

\caption{Number of still alive realizations as a function of time. The inset
indicates an exponential asymptotic decay for GS, after a sudden drop at
$t_{\rm o} \approx 1000$.}

\end{figure}

	Figure~7 shows the number of still alive realizations as a function of
time, again within separated statistics for GS and ST. One observes that some
GS realizations are already dead when the first death among ST occurs. Then,
within a narrow time interval, {\it all}$\,$ ST die. On the other hand, after a
sudden but not extinguishing drop around $t_{\rm o} \approx 1000$, GS
realizations die within a slower rate: some of them survive much more time. The
inset shows this last regime for GS, with linear horizontal scale, indicating
an exponential decay. The characteristic time when the sudden drop occurs
($t_{\rm o} \approx 1000$ in Figure~7) depends on the lattice size $L$, but not
on the number $D$ of dynamics: the larger the lattice size, the later the
system enters into the final exponential decay for GS realizations (inset of
Figure~7). This regime corresponds to a big sea of $up$ spins with some
shrinking islands of neighboring $down$ spins, or vice-versa. It begins at
$t_{\rm o}$, when the spontaneous symmetry breaking finally occurs and one of
the two possible spin orientations up or down has a majority for the first
time: from $t_{\rm o}$ on, this majority fraction increases exponentially fast.
An interesting observation is the coincidence of the beginning of this regime
with the sudden death of all ST realizations, reinforcing once more our
interpretation of ST as mere finite size artifacts: the further exponential
decay is aborted for ST realizations, because the minority islands (narrowest
stripes) are artificially made stable by the boundary conditions.

\begin{figure}[!ht]

\begin{center}
\includegraphics[angle=-90,scale=0.45]{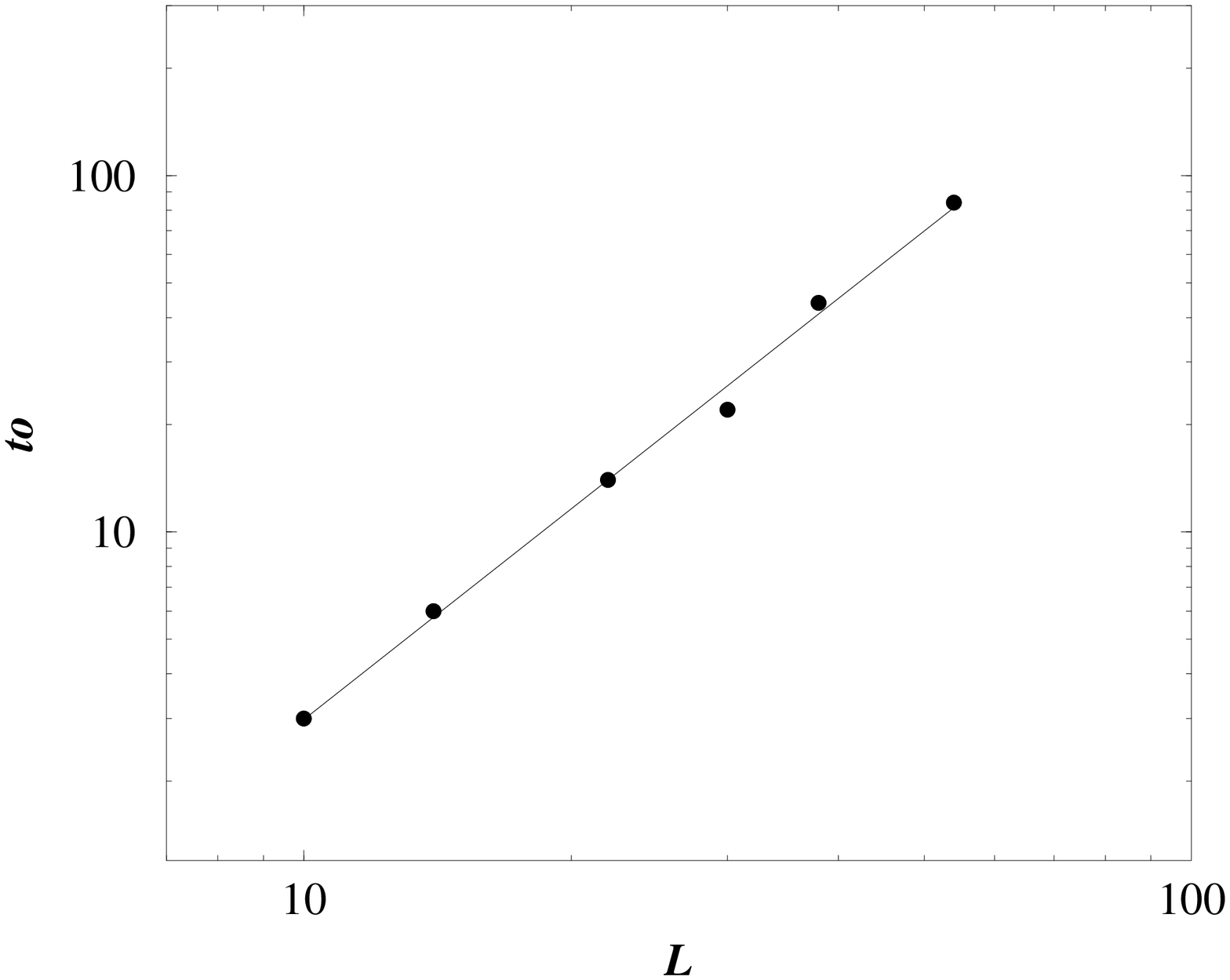}
\end{center}

\caption{Time of first death as function of the lattice size, equation (7).}

\end{figure}

	The characteristic time $t_{\rm o}(L)$ measures the average lifetime
for this evolving system. In order to identify its behavior, in the
thermodynamic limit $L \to \infty$, we measured for each $L = 10$, 14, 22,
30, 38 and 54 the time when the first death occurs among all GS
realizations, adopting $S = 320$ samples with $D = 6000$ dynamics each. The
result is a power-law

\begin{equation}
t_{\rm o}(L)\,\, \propto\,\, L^\alpha\,\,\,\,\,\,\,\, ,
\end{equation}

\noindent with $\alpha \approx 2$, figure~8. A simple reasoning shows the
compatibility of this behavior with that corresponding to persistence, as in
Figure~3. At $t = 1$ (one complete lattice sweep) the average fraction of
non-flipped spins is a constant (numerically 0.708; note also the abrupt
drop from $R(0)=1$ to $R(1)\approx 0.25$ in Figure~1), while at the
characteristic time $t_{\rm o}$ the final value $f$ is reached (Figure~3).
Thus, from equation (5) we can express the corresponding exponent as

\begin{equation}
\theta \,\, =\,\, {\ln({\rm const}) - \ln(f)\over\ln(t_{\rm
o}) - \ln(1)} \,\,\,\,\,\,\,\, {\rm or}\,\,\,\,\,\,\,\, t_{\rm o} \,\,
\propto\,\, f^{-1/\theta}\,\,\,\,\,\,\,\, .
\end{equation}

\noindent Finally, from equation (6) we get

\begin{equation}
t_{\rm o}(L)\,\, \propto\,\, L^{\beta/\theta}\,\,\,\,\,\,\,\, .
\end{equation}

\noindent By comparing equation (7) with (9), we get the scaling relation

\begin{equation}
\alpha\,\, =\,\, \beta/\theta\,\,\,\,\,\,\,\, ,
\end{equation}

\noindent in agreement with our numerical values $\theta = 0.238 \pm
0.002$, $\beta = 0.466 \pm 0.002$ and $\alpha = 1.96 \pm 0.06$.

	An interesting interpretation for the exponent $\alpha = 2$
follows. One particular cluster of neighboring parallel spins grows like
a diffusive random walk, thus with diameter proportional to $t^{1/2}$.
This cluster eventually covers the entire lattice, i.e. $L \propto
t_{\rm o}^{1/2}$. Indeed, by following the growth process of the largest 
cluster just before covering the entire lattice, one observes a typical 
diffusive process.

\section{Conclusions}

	We have studied the dynamical evolution of $2D$ Ising ferromagnets
to explore the extent to which information contained in the randomly
generated initial state determines large-time behavior.  We did this by
comparing different realizations of the dynamical evolution, all starting
from the same initial state, i.e., by monitoring the correlations between
possible alternative different histories, as functions of time. 

	Among other findings, we detected two different regimes during the
time evolution towards the ground state, by counting how many realizations
have already reached it as time goes by. We dicovered the size dependence of
the characteristic relaxation time, $t_{\rm o} \propto L^{\beta/\theta}$,
where $\theta$ is the Derrida exponent and $\beta$ measures the size scaling
of the saturated persistence $f$ (cf.~Eq.~(6) and Figure~3).

	Our most intriguing finding is the suggestion from Figure~6 that the
predictability measure $R_L^{GS}$ may not vanish in the limit $L\to\infty$
so that even in the infinite volume limit there may be predictability of
information about the arbitrarily large time behavior of the system
contained in a randomly generated initial state.  This will be pursued in a
future paper.

\bigskip

Aknowledgements: This work is partially supported by Brazilian agencies
FAPERJ and CNPq (process PRONEX-CNPq-FAPERJ/171.168-2003), and by the
U.S.~National Science Foundation under Grants DMS-01-02587~(CMN) and
DMS-01-02541~(DLS). We thank the referees for several useful comments.

\end{document}